\pdfoutput=1

\documentclass[11pt,twoside,a4paper,cmspaper,final,collab]{cms-tdr}
\def\svnVersion{316882}\def\cmsCernNo{2013-037}\def\cmsCernDate{\today}\def\cmsMessage{Published in Physics Letters B as \href{http://dx.doi.org/10.1016/j.physletb.2015.12.039}{\doi{10.1016/j.physletb.2015.12.039.}}}

\begin{document}\cmsNoteHeader{HIG-14-003}

\hyphenation{had-ron-i-za-tion}
\hyphenation{cal-or-i-me-ter}
\hyphenation{de-vices}
\RCS$Revision: 316085 $
\RCS$HeadURL: svn+ssh://svn.cern.ch/reps/tdr2/papers/HIG-14-003/trunk/HIG-14-003.tex $
\RCS$Id: HIG-14-003.tex 316085 2015-12-24 17:08:28Z andrey $
\newlength\cmsFigWidth
\ifthenelse{\boolean{cms@external}}{\setlength\cmsFigWidth{0.95\columnwidth}}{\setlength\cmsFigWidth{0.45\textwidth}}
\newlength\cmsFigWidthX
\ifthenelse{\boolean{cms@external}}{\setlength\cmsFigWidthX{0.95\columnwidth}}{\setlength\cmsFigWidthX{0.7\textwidth}}
\ifthenelse{\boolean{cms@external}}{\providecommand{\cmsLeft}{top}\xspace}{\providecommand{\cmsLeft}{left}\xspace}
\ifthenelse{\boolean{cms@external}}{\providecommand{\cmsRight}{bottom}\xspace}{\providecommand{\cmsRight}{right}\xspace}
\ifthenelse{\boolean{cms@external}}{\providecommand{\cmsMiddle}{middle}\xspace}{\providecommand{\cmsMiddle}{upper right}\xspace}
\ifthenelse{\boolean{cms@external}}{\providecommand{\cmsUpper}{top}\xspace}{\providecommand{\cmsUpper}{Upper left}\xspace}
\cmsNoteHeader{HIG-14-003}
\title{Search for a Higgs boson decaying into $\gamma^*\gamma\to\ell\ell\gamma$ with low
  dilepton mass in pp collisions at $\sqrt{s}$ = 8\TeV}

\date{\today}

\abstract{A search is described for a Higgs boson decaying into two photons, one of which
  has an internal conversion to a muon or an electron pair ($\ell\ell\gamma$).  The
  analysis is performed using proton-proton collision data recorded with the CMS detector
  at the LHC at a centre-of-mass energy of 8\TeV, corresponding to an integrated
  luminosity of 19.7\fbinv.  The events selected have an opposite-sign muon or electron
  pair and a high transverse momentum photon.  No excess above background has been found
  in the three-body invariant mass range $120<m_{\ell\ell\gamma}<150\GeV$, and limits
  have been derived for the Higgs boson production cross section times branching fraction
  for the decay $\PH\to\gamma^*\gamma\to\ell\ell\gamma$, where the dilepton invariant mass
  is less than 20\GeV.  For a Higgs boson with $m_\PH=125\GeV$, a 95\% confidence level
  (CL) exclusion observed (expected) limit is 6.7 ($5.9^{+2.8}_{-1.8}$) times the standard
  model prediction.  Additionally, an upper limit at 95\% CL on the branching fraction of
  $\PH\to(\JPsi)\gamma$ for the 125\GeV Higgs boson is set at $1.5\times10^{-3}$.}

\hypersetup{%
  pdfauthor={CMS Collaboration},%
  pdftitle={Search for a Higgs boson decaying into gamma* gamma to ll gamma with low
    dilepton mass in pp collisions at sqrt(s) = 8 TeV },%
  pdfsubject={CMS},%
  pdfkeywords={CMS, Physics, Higgs boson, Dalitz decay}}

\maketitle
\section{Introduction}
\label{sec:intro}

\begin{figure*}[bht]
  \centering
    {\includegraphics[width=0.23\textwidth]{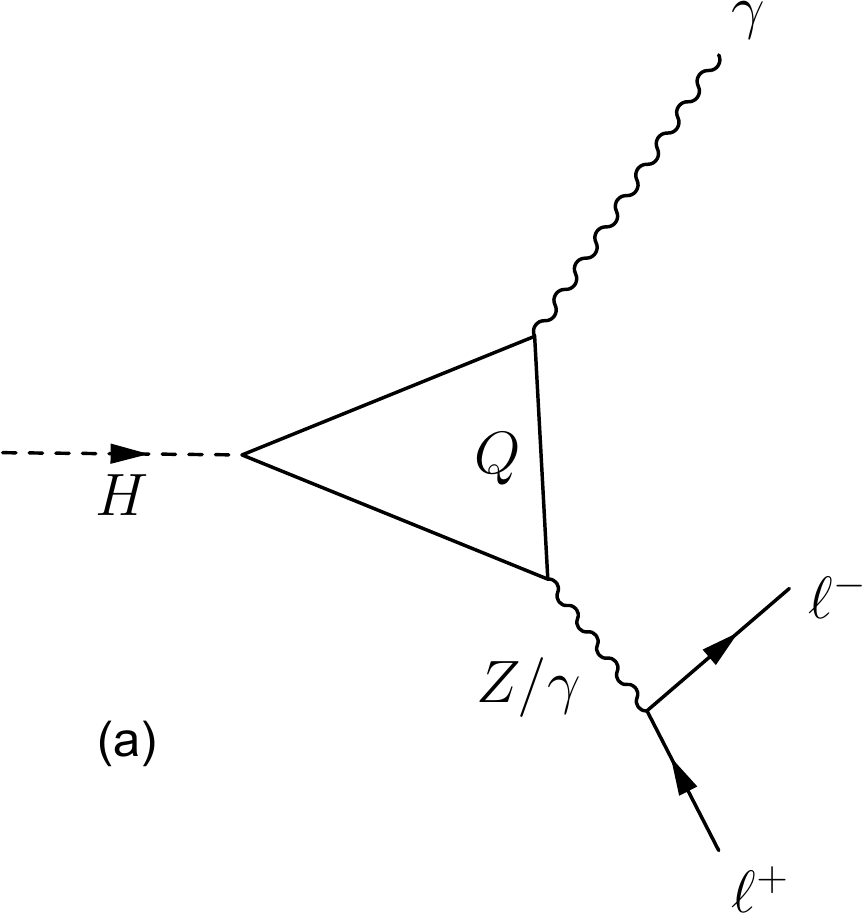}}\hfil
    {\includegraphics[width=0.23\textwidth]{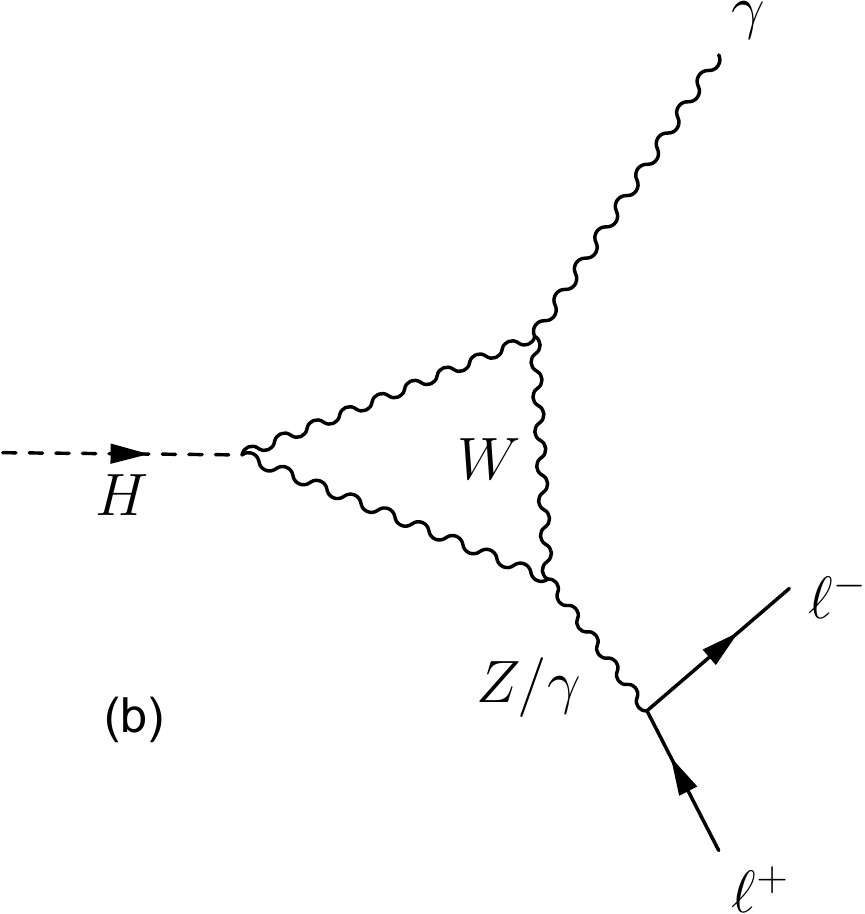}}\hfil
    {\includegraphics[width=0.23\textwidth]{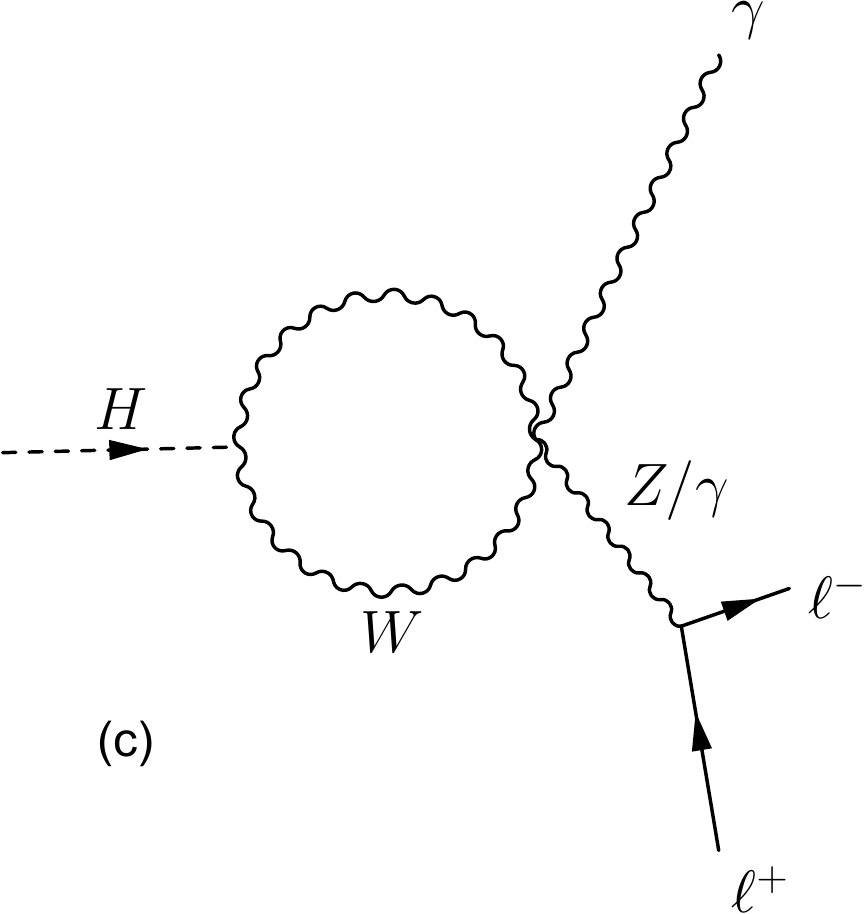}}
    {\includegraphics[width=0.23\textwidth]{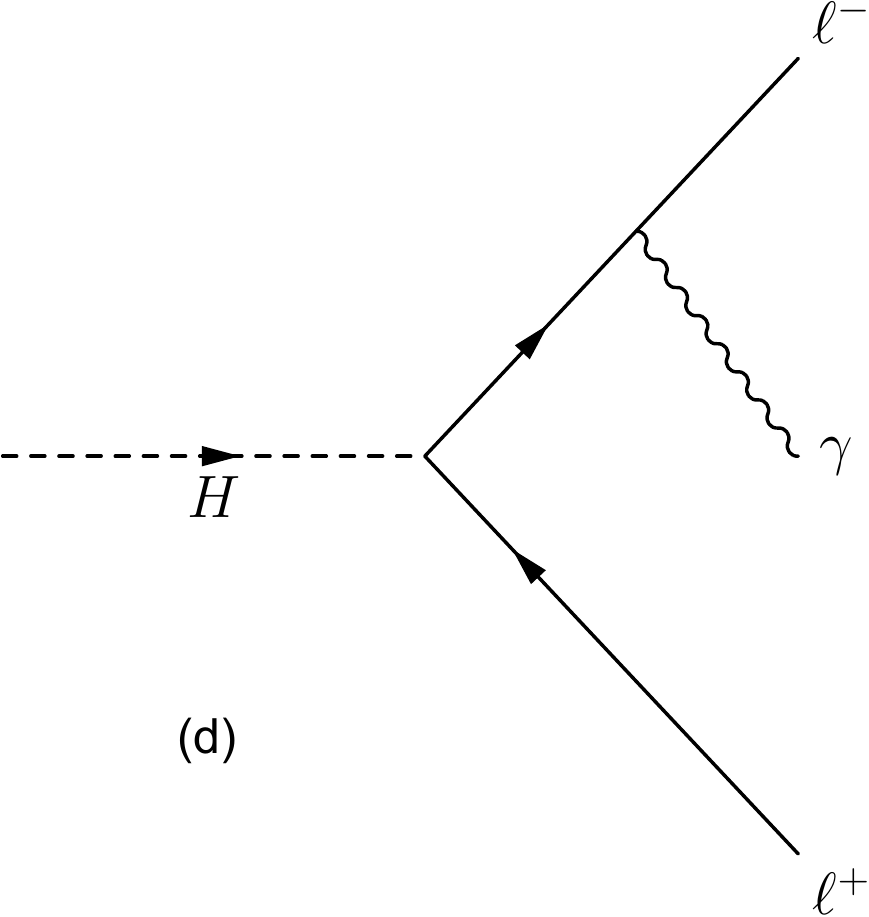}} \hfil
    {\includegraphics[width=0.23\textwidth]{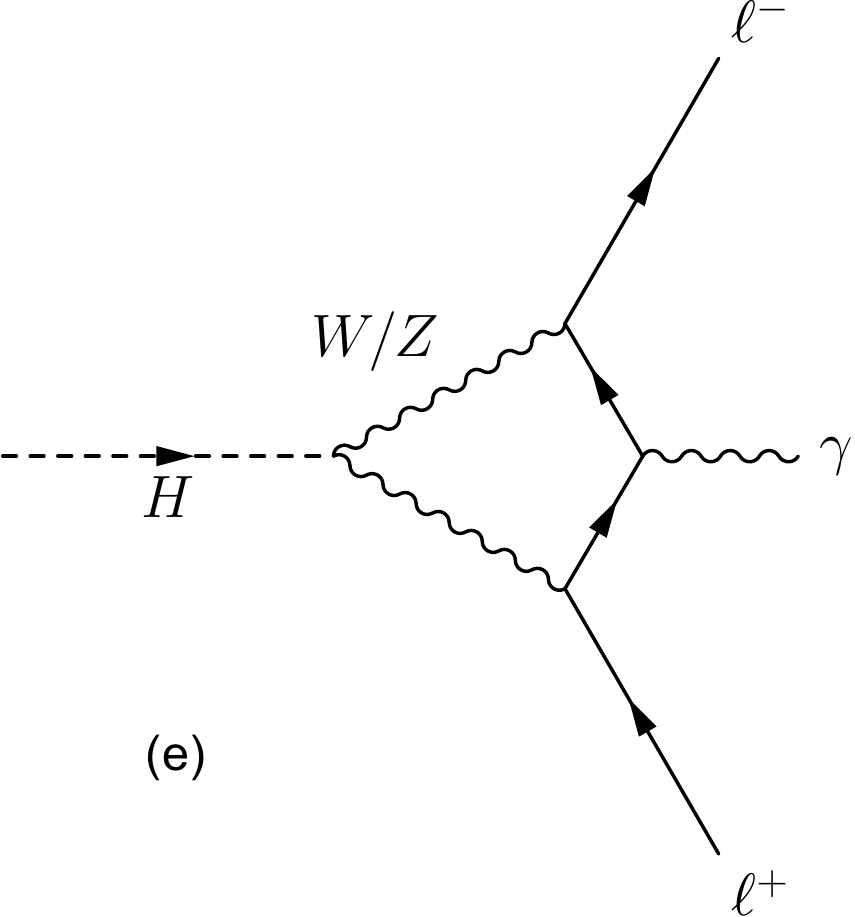}} \hfil
    {\includegraphics[width=0.23\textwidth]{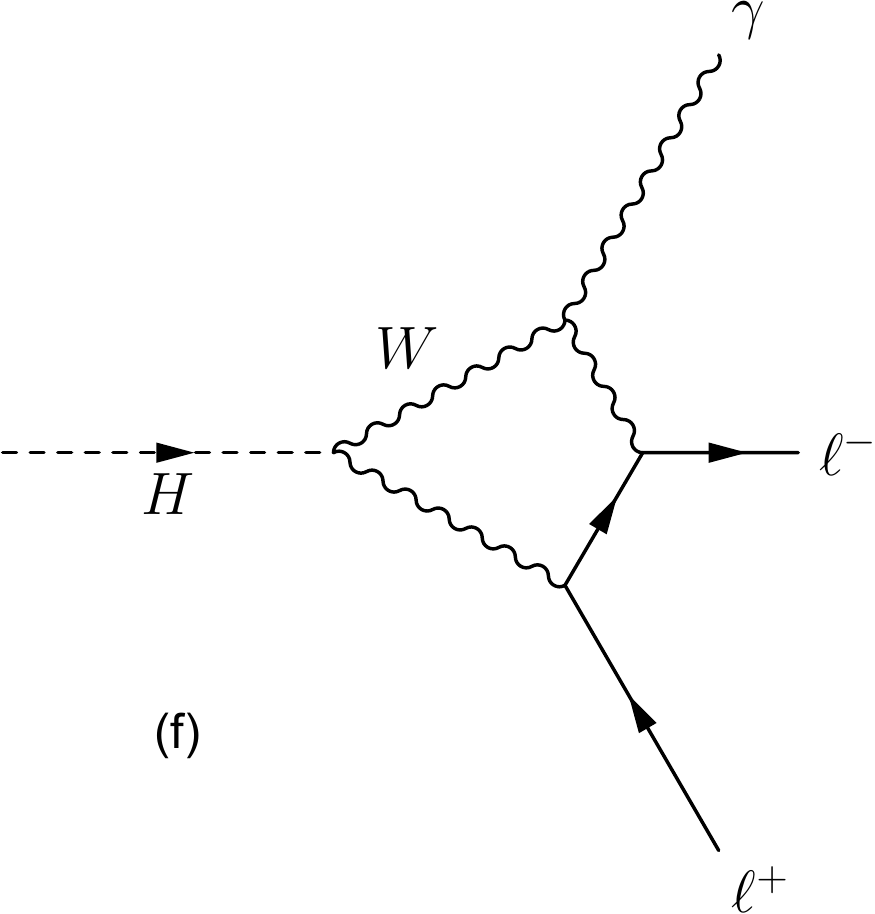}} \hfil
    {\includegraphics[width=0.23\textwidth]{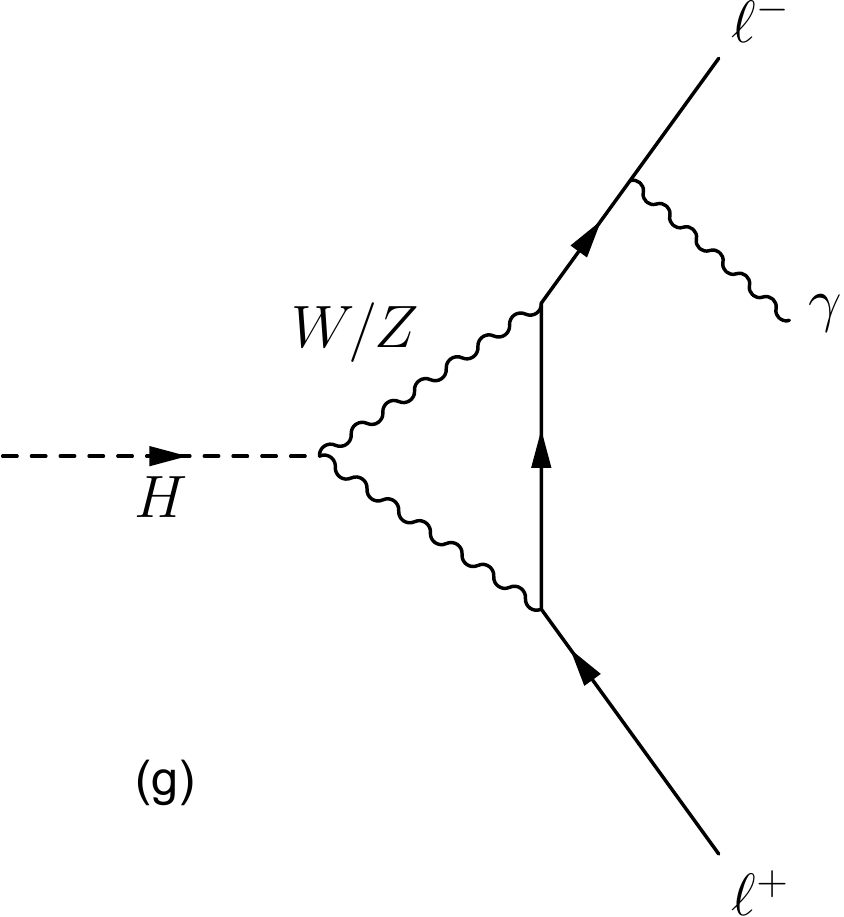}}
  \caption{Diagrams contributing to $\PH \to \ell\ell\gamma$. The contributions from
    diagrams (a), (b), and (c) dominate.  The final-state radiation of $\PH\to\mu\mu$
    decay, shown in diagram (d), is important at high dilepton invariant mass.  Higher
    order contributions from diagrams (e), (f) and (g) are negligible.  }
  \label{fig:diag}

\end{figure*}

\begin{figure*}[thb]
  \centering\hfil
  \includegraphics[width=0.35\textwidth]{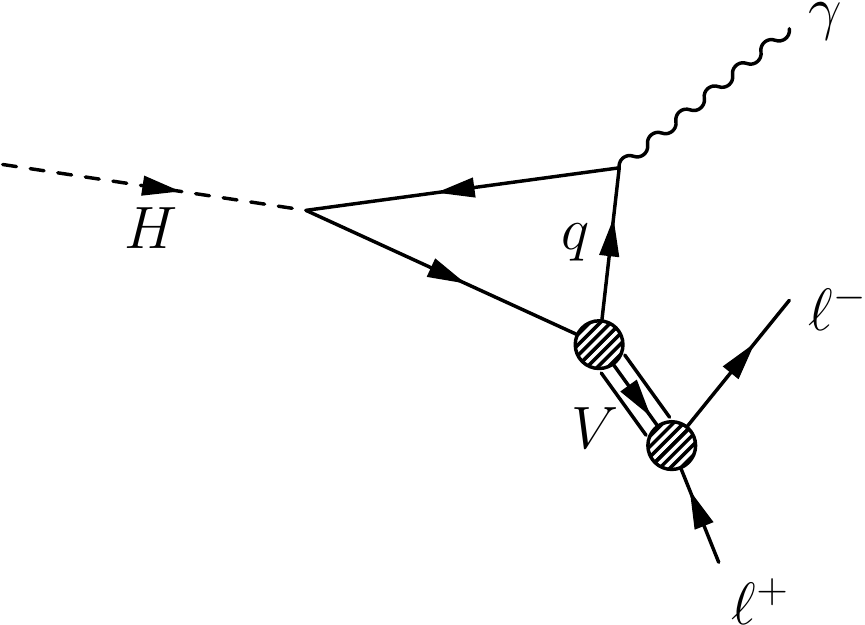}\hfil
  \includegraphics[width=0.35\textwidth]{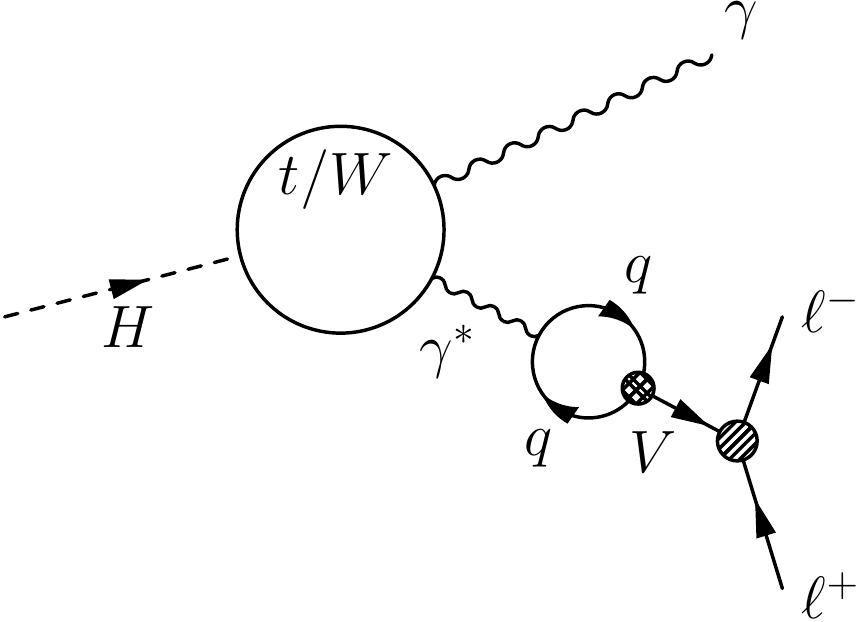}\hfil
  \caption{Diagrams contributing to $\PH \to V\gamma \to \ell\ell\gamma$ decay.}
  \label{fig:hjp}

\end{figure*}

The rare decay into the $\ell\ell\gamma$ final state of the Higgs boson is a rich source
of information that can enhance our understanding of its basic properties and probe novel
couplings predicted by extensions of the standard model\,(SM) of particle physics.  As
illustrated in Fig.~\ref{fig:diag}, this decay in SM has contributions from loop-induced
$\PH\to\gamma^*\gamma$ and $\PH \to \Z\gamma$ diagrams (a, b, c), tree-level process
$\PH\to\ell\ell$ with final-state radiation (d), and higher-order processes, known as box
diagrams (e, f, g)~\cite{Abba96,Chen12,Dicus13,Passarino}.  Other contributions include
$\PH \to V(q\bar{q})\gamma \to \ell\ell\gamma$, shown in Fig.~\ref{fig:hjp}, where $V$
denotes a vector meson ($\JPsi$ or $\Upsilon$) that decays to $\ell\ell$
~\cite{hToJPsiGamma-2013, hToJPsiGamma-2014, Koenig}.  The Higgs boson branching fraction
to $\ell\ell\gamma$ is dominated by the $\PH\to\gamma^*\gamma$ and $\PH \to \Z\gamma$
modes, while the contribution from the box diagrams is negligible~\cite{Abba96}.  In the
muon channel, when the dilepton invariant mass, $m_{\ell\ell}$, is greater than 100\GeV,
final-state radiation in $\PH\to\mu\mu$ starts to dominate~\cite{Htollg-FB-Sun}.  In the
three-body decay, $\PH\to\ell\ell\gamma$, it is possible to investigate non-SM couplings
by examining the angular distributions, and forward-backward asymmetry variables
reconstructed from the $\ell\ell\gamma$ final state~\cite{Htollg-FB-Sun, Htollg-FB-Kor}.

The expected rates of the $\PH\to (\Z/\gamma^*)\gamma\to \ell\ell\gamma$ processes
compared to the rate of $\PH\to\gamma\gamma$ decay, for a Higgs boson with mass
$m_\PH=125\GeV$, are~\cite{Firan07,Heinemeyer:2013tqa}:

\ifthenelse{\boolean{cms@external}}{
\begin{equation*}\begin{split}
    \frac{\Gamma(\PH\to\gamma^*\gamma \to \Pe\Pe\gamma)}{\Gamma(\PH\to\gamma\gamma)} &\sim 3.5\%, \\
    \frac{\Gamma(\PH\to\gamma^*\gamma \to \mu\mu\gamma)}{\Gamma(\PH\to\gamma\gamma)} &\sim 1.7\%, \\
    \frac{\Gamma(\PH\to\Z\gamma\to \ell\ell\gamma)}{\Gamma(\PH\to\gamma\gamma)} &\sim 2.3\%.
\end{split}\end{equation*}
}{
\small
\begin{equation*}
    \frac{\Gamma(\PH\to\gamma^*\gamma \to \Pe\Pe\gamma)}{\Gamma(\PH\to\gamma\gamma)} \sim 3.5\%, \quad
    \frac{\Gamma(\PH\to\gamma^*\gamma \to \mu\mu\gamma)}{\Gamma(\PH\to\gamma\gamma)} \sim 1.7\%, \quad
    \frac{\Gamma(\PH\to\Z\gamma\to \ell\ell\gamma)}{\Gamma(\PH\to\gamma\gamma)} \sim 2.3\%.
\end{equation*}
\normalsize
}

The $\PH\to\gamma^*\gamma\to \Pe\Pe\gamma$ decay is distinct from $\PH\to\gamma\gamma$
followed by a conversion of a photon to an $\Pep\Pem$ pair in the detector, which can
become a background for $\PH\to\gamma^*\gamma$ if photon conversions are not properly
identified.  Experimentally, the various contributions shown in Figs.~\ref{fig:diag}
and \ref{fig:hjp} can be disentangled to some extent.  Requirements on $m_{\ell\ell}$ and
the transverse momentum (\PT) of the photon are used to separate $\PH\to\gamma^*\gamma$
and $\PH\to\Z\gamma$. Events with final-state radiation are removed by requiring the
photon to be isolated from either of the leptons.  Contributions from $\PH \to
(\JPsi)\gamma \to
\ell\ell\gamma$ and other resonances are identified and rejected or selected based on the
value of $m_{\ell\ell}$.

The ATLAS and CMS Collaborations at the CERN LHC have both performed a search for $\PH \to
\Z\gamma\to\ell\ell\gamma$ decay with $m_{\ell\ell}$ above 50\GeV~\cite{atl-HZG, cms-HZG}.
As a natural extension of those analyses, the current paper describes the first search for
a Higgs boson Dalitz decay, $\PH\rightarrow\gamma^*\gamma$, where the $\gamma^*$ decays
into a muon or an electron pair.  The search is performed for a Higgs-like particle within
the mass range between 120 and 150\GeV.  In order to select the contribution from the
Dalitz decay, we require $m_{\ell\ell} < 20\GeV$.  The $\mu\mu\gamma$ topology is a clean
final state with a mass resolution of about 1.6\%, as measured from the simulated signal
samples.  The $\Pe\Pe\gamma$ channel is challenging due to the low $m_{\ell\ell}$ that
results in a pair of merged electron showers in the electromagnetic calorimeter (ECAL).
Nevertheless, when the merged showers are reconstructed in the ECAL, a mass resolution of
1.8\% is achieved.  Important backgrounds include the irreducible contributions from the
initial- and final-state photon radiation in Drell--Yan production, and Drell--Yan events
with additional jets where a jet is misidentified as a photon.

In addition, a search is performed for $\PH \to (\JPsi)\gamma\to\mu\mu\gamma$ decay for
$m_\PH=125\GeV$, which is sensitive to the Higgs boson coupling to charm quark and a
promising way to access the couplings of the Higgs boson to the second generation quarks
at the LHC.  In the SM this decay occurs through two main processes: direct coupling of
the Higgs boson to charm (Fig.~\ref{fig:hjp}a), and the usual t/W loop, where the radiated
$\gamma^*$ converts to a $c\bar{c}$ in a resonant state (Fig.~\ref{fig:hjp}b). The two
amplitudes interfere destructively and the second one
dominates~\cite{hToJPsiGamma-2014,Koenig}. For the SM Higgs boson with $m_\PH=125\GeV$,
the branching fraction is predicted to be $2.8\times10^{-6}$.  A search by the ATLAS
Collaboration for this decay is described in Ref.~\cite{ATL-hjp}.

The results presented in this paper are based on proton-proton collision data recorded in
2012 with the CMS detector at a centre-of-mass energy $\sqrt{s}=8\TeV$, corresponding to
an integrated luminosity of 19.7\fbinv.

\section{CMS detector and trigger}
\label{sec:cms}
A detailed description of the CMS detector, together with a definition of the coordinate
system used and the relevant kinematic variables, can be found in Ref.~\cite{ref:cms}.
The central feature of the CMS apparatus is a superconducting solenoid, 13\unit{m} in
length and 6\unit{m} in diameter, which provides an axial magnetic field of 3.8\unit{T}.
Within the solenoid volume are a silicon pixel and strip tracker, the ECAL, and a hadron
calorimeter (HCAL).  Charged-particle trajectories are measured by silicon pixel and strip
trackers, covering $0\le \phi \le 2\pi$ in azimuth and $\abs{\eta} < 2.5$ in
pseudorapidity. A lead tungstate crystal ECAL surrounds the tracking volume. It is
comprised of a barrel region $\abs{\eta} < 1.48$ and two endcaps that extend up to
$\abs{\eta} = 3$. A brass and scintillator HCAL surrounds ECAL and also covers the region
$\abs{\eta} < 3$.  Iron forward calorimeters with quartz fibers, read out by
photomultipliers, extend the calorimetric coverage up to $\abs{\eta} = 5$.  A lead and
silicon-strip preshower detector is located in front of the ECAL endcaps. Muons are
identified and measured in gas-ionization detectors embedded in the steel flux-return yoke
outside the solenoid. The detector is nearly hermetic, allowing energy balance
measurements in the plane transverse to the beam direction.

A two-tier trigger system selects collision events of interest for physics analysis. Two
triggers are used in the current analysis.  In the muon channel, the trigger requires a
single muon and a photon, both with \pt greater than 22\GeV.  In the electron channel the
$\gamma^* \to \Pe\Pe$ process at low dielectron invariant mass mimics a photon at the
trigger level.  For this reason, a diphoton trigger is used in the electron channel, for
$\gamma + \gamma^*$ final state.  The trigger requires a leading (subleading) photon with
$\PT > 26\,(18)\GeV$.  The diphoton trigger is inefficient for events with high dielectron
invariant mass ($m_{\Pe\Pe} > 2\GeV$) due to the isolation and shower shape
requirements. The available dielectron triggers cannot be used to select events with $2 <
m_{\Pe\Pe} < 20\GeV$ because they also require isolation, and the \PT requirement made
on the subleading lepton is too high.

\section{Event reconstruction}
\label{sec:reco}
The photon energy is reconstructed from a sum of signals in the ECAL
crystals~\cite{EGM-14-001}.  The ECAL signals are calibrated and
corrected~\cite{Calib-ECAL}, and a multivariate regression technique, developed for the
$\PH\to\gamma\gamma$ analysis~\cite{cms-Hgg-Legacy}, is used to determine the final energy
of the photon~\cite{EGM-14-001}.  The neighboring ECAL crystals with energy deposition are
combined into clusters, and the collection of clusters that contain the energy of a photon
or an electron is called a supercluster.  Identification criteria are applied to
distinguish photons from jets and electrons.  The observables used in the photon
identification criteria are: the isolation variables, the ratio of the energy in the HCAL
towers behind the supercluster to the electromagnetic energy in the supercluster; the
transverse width in $\eta$ of the electromagnetic shower; and the number of charged tracks
matched to the supercluster.  The efficiency of the photon identification is measured
using $\Z \to \Pe\Pe$ data by reconstructing the electron showers as photons, and found to
be 80\,(88\%) at a transverse energy $>30\,(50)\GeV$ and $\abs{\eta}<1.44$.

\begin{figure}[t]
  \centering

    \includegraphics[width=\cmsFigWidth]{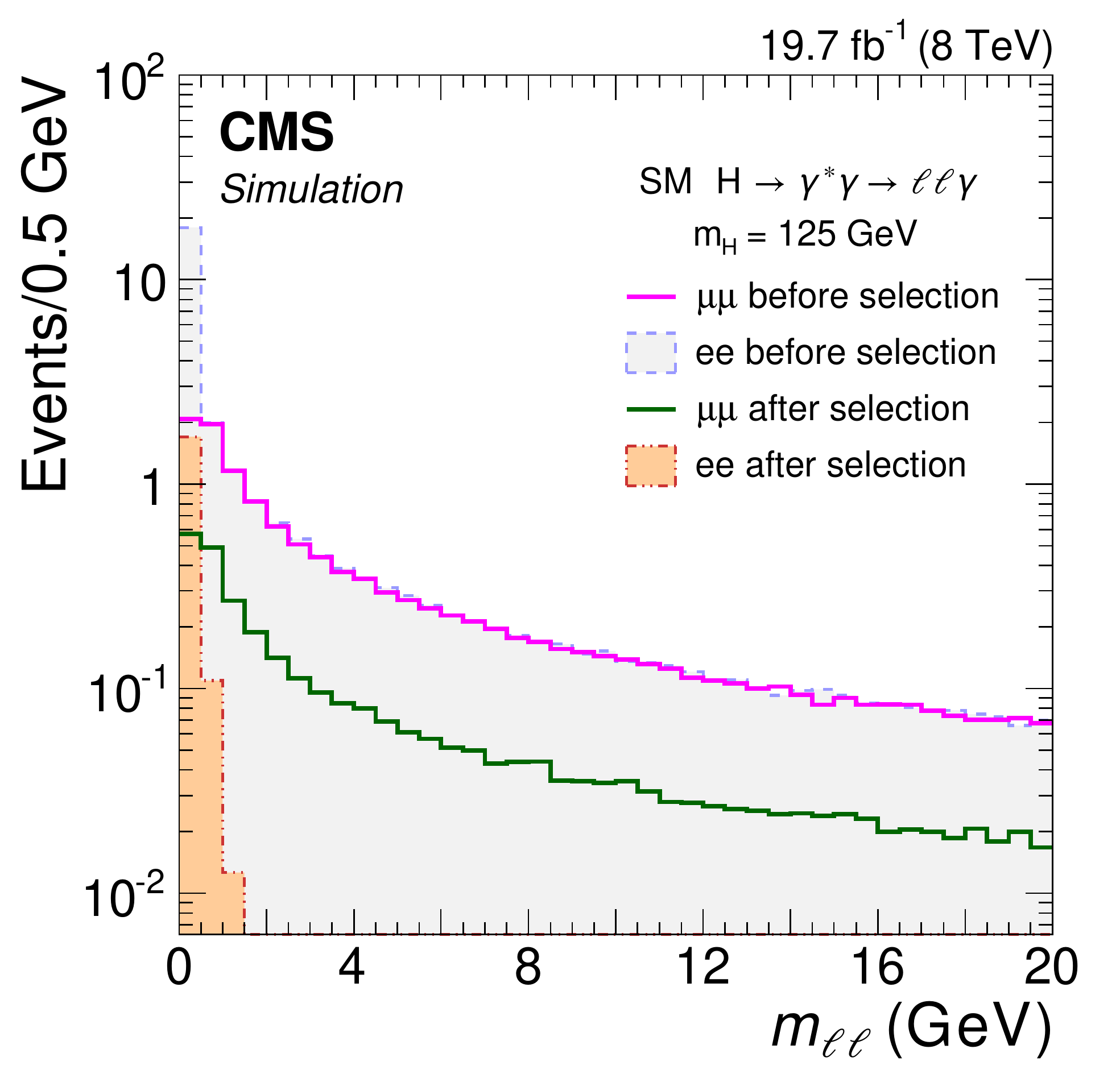}
    \caption{The invariant mass of the dilepton system
      in signal simulation for $m_\PH=125\GeV$.
      Distributions are shown for muon and electron channels, before and after selection.
      The invariant mass before selection is obtained from the leptons at the generator level,
      while after selection the reconstructed invariant mass is used.}
    \label{fig:mll}

\end{figure}

Muon candidates are reconstructed in the tracker and identified by the particle-flow
global event reconstruction algorithm~\cite{PF09,PF10} using hits in the tracker and the
muon systems.  This approach allows us to maintain a high efficiency independent of the
dimuon invariant mass and to reconstruct muons with \PT as low as 4\GeV.  Muons from
$\gamma^*\to\mu\mu$ internal conversions are expected to be isolated from other particles.
A cone of size $\DR \equiv\sqrt{(\Delta\eta)^2 + (\Delta\phi)^2} = 0.4$ is constructed
around the momentum direction of each muon candidate~\cite{CMS-muon}.  The relative
isolation of the muon is quantified by summing the \PT of all photons, charged and neutral
hadrons within this cone, and then dividing by the muon \PT.  The resulting quantity,
corrected for additional underlying event activity due to pileup events, is required to be
less than 0.4 for the leading muon.  The isolation requirement rejects misidentified
leptons and background arising from hadronic jets.  The $\DR(\mu\mu)$ separation between
the two muons is small due to their low invariant mass (as shown in Fig.~\ref{fig:mll})
and high $\PT$ of the $\gamma^*$ in $\PH\to\gamma^*\gamma$ decays.  Hence, no isolation
requirement is applied to the subleading muons as they are already within the isolation
cone of the leading muons in most events.  Dimuon identification and isolation efficiency
of about 80\% is obtained.

In the electron channel of the $\PH\to\gamma^*\gamma\to \ell\ell\gamma$ decay, the two
electrons produced in the $\gamma^*\to\Pe\Pe$ process are even closer to each other than
in the muon channel, since the $m_{\ell\ell}$ is smaller~(Fig.~\ref{fig:mll}).  Therefore,
their energy deposits in the ECAL are merged into one supercluster giving rise to a unique
signature.  To identify these merged electrons, two tracks associated to the supercluster
are required. A Gaussian sum filter (GSF) algorithm is used to reconstruct the electron
tracks~\cite{ele-GSF}.  The supercluster energy must correspond to $\PT>30\GeV$ and be
located in the ECAL barrel ($\abs{\eta}<1.44$). The scalar sum $\PT^{\Pe_1}+ \PT^{\Pe_2}$
of the corresponding two GSF tracks must exceed $44\GeV$.  Both GSF tracks are required to
have no more than one missing hit in the pixel detector in order to reduce the background
from photons converting to $\Pep\Pem$ in the detector material.  A multivariate
discriminator is trained to separate the $\gamma^*\to\Pe\Pe$ objects from jets or single
electrons.  The input variables for the training include lateral shower shape variables,
the median energy density in the event to take into account the pileup dependence, and the
kinematic information from the supercluster and tracks.  A combined reconstruction and
selection efficiency of ${\sim}40\%$ is achieved for the signal. For comparison, the
efficiency for a single isolated electron with similar \PT is
${\sim}88\%$~\cite{EGM-13-001}.

\section{Simulated samples}
\label{sec:sim}

The description of the Higgs boson signal used in the search is obtained from simulated
events.  The samples for the Dalitz signal are produced at leading-order using the
\MADGRAPH\,5 matrix-element generator~\cite{MAD5} with the ANO-HEFT model~\cite{ANO-HEFT},
interfaced with {\PYTHIA\,6.426} \cite{pythia6}, for the gluon and vector boson fusion
processes, and for associated production with a vector boson.  Associated production with
a \ttbar pair is ignored because of its small contribution.  The sample for
$\PH\to(\JPsi)\gamma$ process is produced with the {\PYTHIA\,8.153}
generator~\cite{pythia8}, and reweighted to simulate 100\% polarization of the \JPsi.  The
parton distribution function (PDF) set used to produce these samples is given by
CTEQ6L1~\cite{CTEQ6L}.  The SM Higgs boson production cross sections are taken from
Ref.~\cite{Heinemeyer:2013tqa}.  The branching fractions for $\PH\to\gamma^*\gamma$ are estimated using
MCFM 6.6~\cite{MCFM-br} and for $\PH\to(\JPsi)\gamma$ are taken from
Ref.~\cite{hToJPsiGamma-2014}.  For the SM Higgs boson in the mass range of 120--150\GeV,
the $\PH\to\gamma^*\gamma\to\mu\mu\gamma\,(\Pe\Pe\gamma)$ branching fraction is expected
to be between $2.0\,(4.5) \times 10^{-5}$ and $3.3\,(7.5) \times 10^{-5}$ for
$m_{\ell\ell}$ below 20\GeV.  The expected branching fraction for $\PH \to (\JPsi)\gamma$
is $(2.8\pm0.2)\times10^{-6}$ for $m_\PH=125\GeV$, which is further suppressed due to the
$\JPsi$ meson decay to muons, $\mathcal{B}(\JPsi\to\mu\mu) = 0.059$.

The simulation aims to include all known effects and the conditions of real data taking in
CMS. Some residual differences between the data and simulation are taken into account by
reweighting the simulated events with scale factors. Systematic uncertainties are assigned
to cover imperfect knowledge of residual differences. Scale factors are implemented to
match the distribution of primary vertices, the photon identification and isolation
efficiency, and the muon isolation efficiency. No corrections are applied to the muon and
electron identification and trigger efficiencies, but an uncertainty is assigned as
described in Section~\ref{sec:results}.

The energy and momentum resolution of muons and photons in simulated events are corrected
to match that in data. The energy scale of muons (photons) is corrected to that found in
$\Z\to\mu\mu\,(\Pe\Pe)$ events.  For the electrons, no resolution or scale corrections are
applied because of their unique topology, and the absence of a data-driven method to
derive those corrections.  Therefore, we rely on the simulation of the $\gamma^*\to\Pe\Pe$
process and assign uncertainties sufficient to cover any possible discrepancy in the scale
and resolution between data and simulation.

\section{Event selection}
\label{sec:selection}

Events are required to pass the muon plus photon trigger in the $\mu\mu\gamma$ final state
and the diphoton triggers in the $\Pe\Pe\gamma$ final state.  The trigger efficiency for
signal events after the selection requirements described below is 85\% (90\%) in the muon
(electron) channel, as measured from the simulated samples.

The muons (electrons) are required to be within $\abs{\eta}<2.4\,(1.44)$, while the photon
is required to be within $\abs{\eta}<1.44$.  The invariant mass of the $\ell\ell\gamma$
system, $m_{\ell\ell\gamma}$, is required to satisfy $110 < m_{\ell\ell\gamma} <
170\GeV$. The photon and dilepton momenta both must satisfy $\pt > 0.3\cdot
m_{\ell\ell\gamma}$ requirement, which is optimized for high signal efficiency and
background rejection.

On average, there are 21 $\Pp\Pp$ interactions within the same bunch crossing in the 8\TeV
data, which result in about 16 collision vertices reconstructed in each event. The vertex
with the highest scalar sum of the ${\pt^2}$ of its associated tracks is taken to
correspond to the primary interaction vertex.  The primary vertex must have the
reconstructed longitudinal position ($z$) within 24\unit{cm} of the geometric centre of
the detector and the transverse position ($x$-$y$) within 2\unit{cm} of the beam
interaction region.  The lepton tracks from $\gamma^*\to\mu\mu$ ($\Pe\Pe$) are required to
originate from the primary vertex, and to have transverse and longitudinal impact
parameters with respect to that vertex smaller than $2.0\,(0.2)\unit{mm}$ and
$5\,(1)\unit{mm}$, respectively.

The muons must be oppositely charged, and have $\PT > 23\,(4)\GeV$ for the leading
(subleading) lepton.  The \PT requirement on the leading muon is driven by the trigger
threshold, and on the subleading muon by the minimum energy needed to reach the muon
system, while maintaining high reconstruction efficiency.  In the electron channel, no
additional selection on \PT of the GSF tracks is necessary, beyond those described in
Section~\ref{sec:reco}.  Finally, in both muon and electron channels, the separation
between each lepton and the photon is required to satisfy $\DR>1$ in order to suppress
Drell--Yan background events with final-state radiation.

The dilepton invariant mass in the muon channel is required to be less than 20\GeV to
reject contributions from $\Pp\Pp\to\gamma\Z$ and to suppress interference effects from
the $\PH\to\gamma\Z$ process and the box diagrams shown in Fig.~\ref{fig:diag}.  Events
with a dimuon mass in the ranges $2.9< m_{\mu\mu}< 3.3\GeV$ and $9.3 < m_{\mu\mu} <
9.7\GeV$ are rejected to avoid the $\JPsi\to\mu\mu$ and $\Upsilon\to\mu\mu$
contamination.  In the electron channel the invariant mass, constructed from the two GSF
tracks, is required to satisfy $m_{\Pe\Pe} < 1.5\GeV$.  The $m_{\ell\ell}$ distributions
for simulated signal events are shown in Fig.~\ref{fig:mll} in the muon and electron
channels.

In the search for the $\PH\to(\JPsi)\gamma\to\mu\mu\gamma$, both $\pt^{\gamma}$ and
$\pt^{\mu\mu}>40\GeV$ are required, and the events are selected with $2.9 <
m_{\mu\mu}<3.3\GeV$.

The observed yields after the event selection described above are listed in
Table~\ref{tab:yield}. In the electron channel, there is also a contribution from the
$\PH\to\gamma\gamma$ process due to unidentified conversions, which is about 15\% of the
$\PH\to\gamma^*\gamma$ signal (0.2 events at $m_\PH = 125\GeV$).  This contribution is
considered as a background to $\PH\to\gamma^*\gamma$, and negligible compared to the
continuum background estimated from the fit to data described in the next section.

\begin{figure}[t]
  \centering
    \includegraphics[width=\cmsFigWidth]{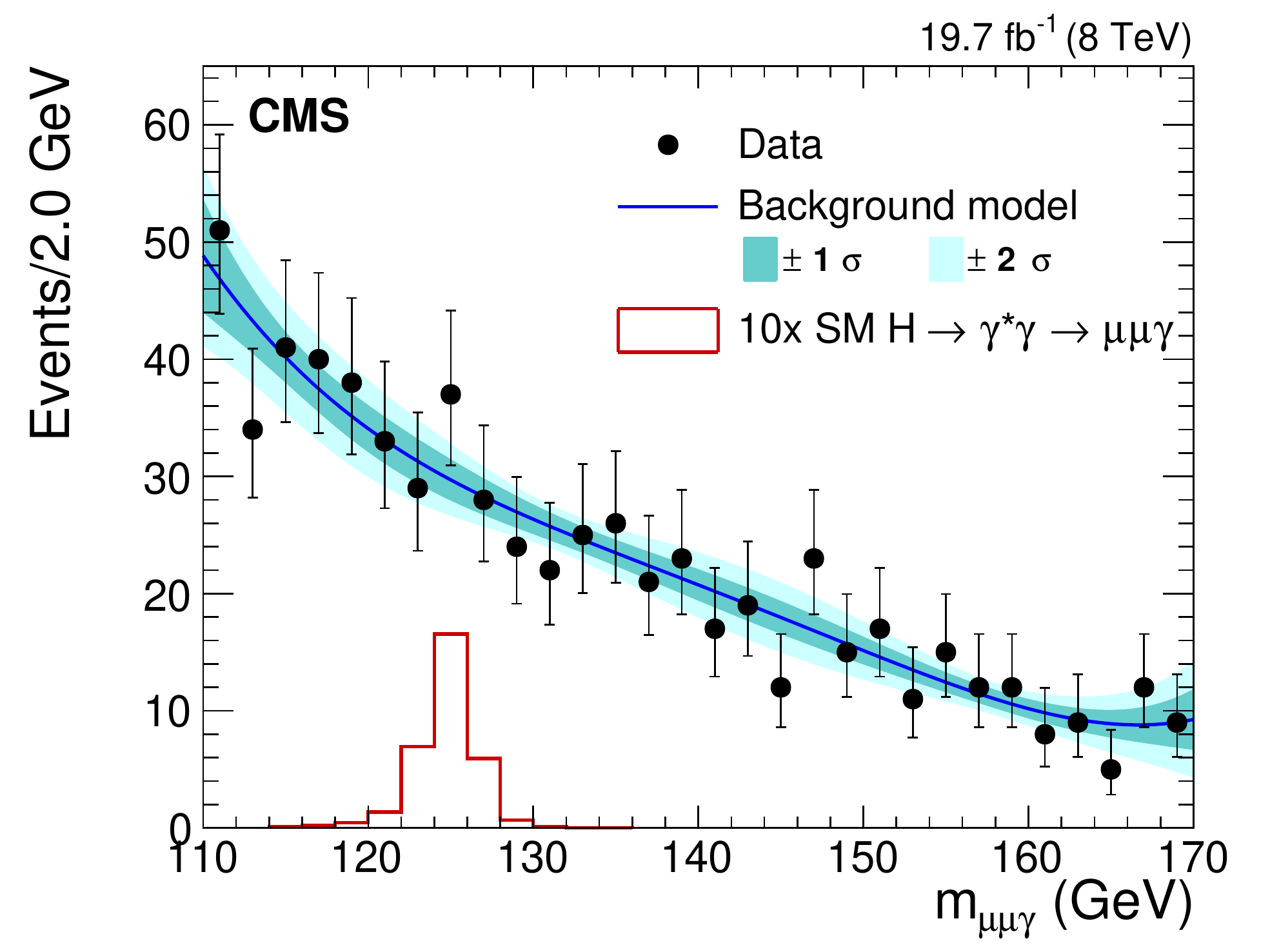}
    \includegraphics[width=\cmsFigWidth]{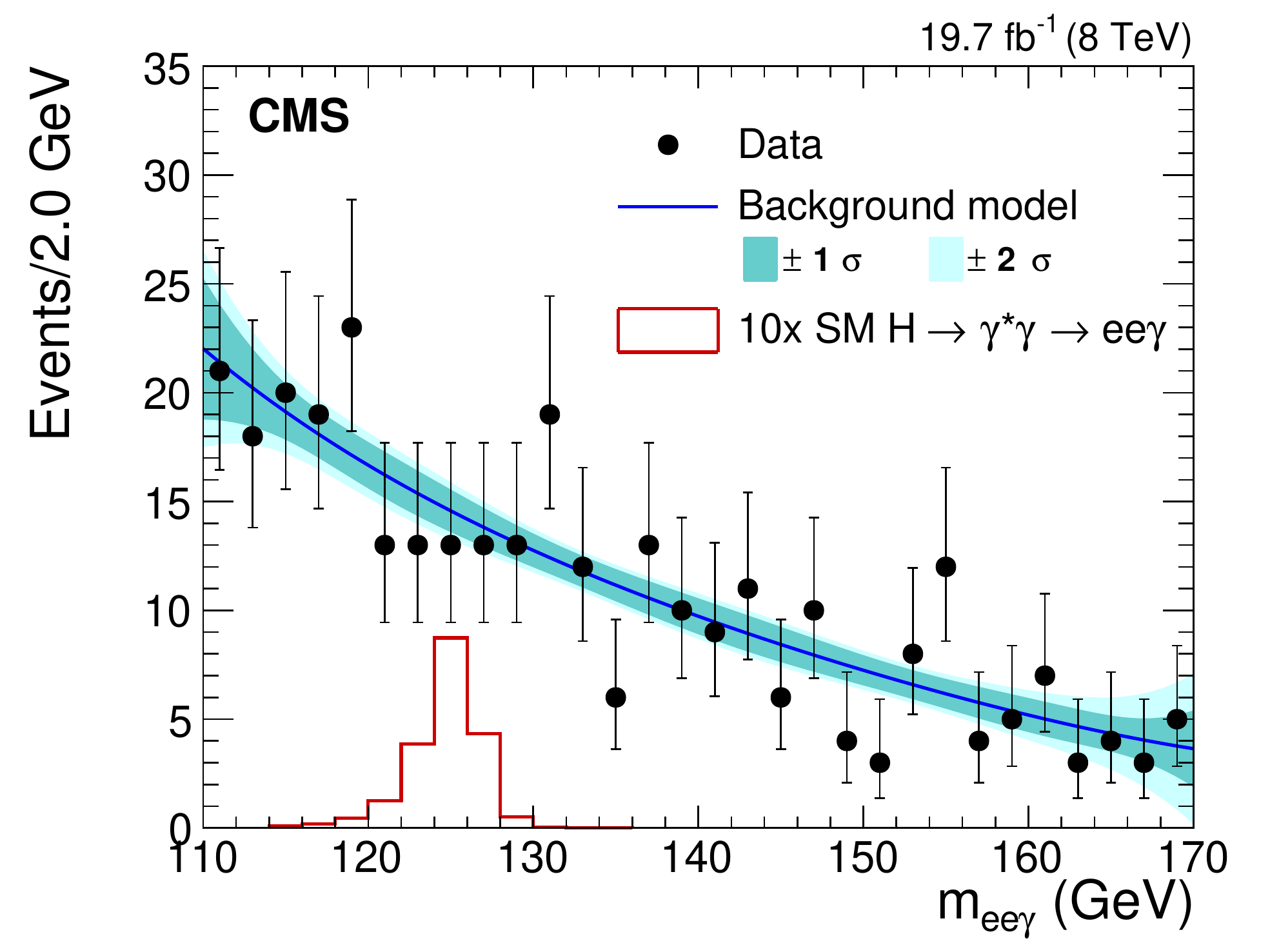}
    \caption{The $m_{\mu\mu\gamma}$ (\cmsLeft) and $m_{\Pe\Pe\gamma}$ (\cmsRight) spectra for
      8\TeV data (points with error bars), together with the result of a background-only
      fit to the data. The 1$\sigma$ and 2$\sigma$ uncertainty bands represent the
      uncertainty in the parameters of the fitted function.
      The expected contribution from the SM Higgs boson
      signal with $m_\PH=125\GeV$, scaled up by a factor of~10, is shown as a histogram.
           \label{fig:fit}
         }
\end{figure}

\begin{table*}[bth]
  \topcaption{The expected signal yield and the number of events in data, for an integrated luminosity of $19.7\fbinv$.
    Signal events are presented before and after applying the full selection criteria described in the text.
    In the $(\JPsi)\gamma$ sub-category only the $\JPsi\to\mu\mu$ decay is considered,
    and the signal yield is a sum of two contributions:
    $\PH\to(\JPsi)\gamma\to\mu\mu\gamma$ and $\PH\to\gamma^*\gamma\to\mu\mu\gamma$, where the dimuon mass distribution is non-resonant.
  }
  \label{tab:yield}
  \centering
  \resizebox{\textwidth}{!}{
      \begin{tabular}{l c c|c}
        \multirow{3}{*}{Sample}&  Signal events         &  Signal events                & Number of events                        \\
        & before selection       & after selection               & in data                              \\
        & $m_\PH=125\GeV$  & $m_\PH=125\GeV$               & $120 < m_{\ell\ell\gamma} <130\GeV$  \\
        \hline
        $\mu\mu\gamma$           & 13.9                       & 3.3                                & 151 \\
        $\Pe\Pe\gamma$           & 25.8                       & 1.9                                & 65 \\

        \hline
        $(\JPsi\to\mu\mu)\gamma$ & $0.065 (\JPsi) +  0.32 \text{ (non-res.)}$ & $0.014 (\JPsi) + 0.078 \text{ (non-res.)}$ & 12 \\
      \end{tabular}
}
\end{table*}

\section{Background and signal modeling}
\label{sec:background_signal}
The background is modeled by fitting a polynomial function to the $\ell\ell\gamma$ mass
distributions in data.  An unbinned maximum likelihood fit is performed over the range
$110<m_{\ell\ell\gamma}<170\GeV$.  Fig.~\ref{fig:fit} shows the $m_{\ell\ell\gamma}$
spectra, which are fitted with polynomial functions of fourth degree.  The reduced
$\chi^2$ of the fits are 0.5 and 0.7 for the muon and electron channels, respectively.
Even though the search is limited to $120<m_\PH<150\GeV$, the fits to the
$m_{\ell\ell\gamma}$ spectra are performed over a wider range, giving a better modeling of
the background, particularly at the edges of the search range.  The degree of the
polynomials is chosen following a procedure similar to the one described in
Ref.~\cite{long}. This procedure ensures that the potential bias due to the background
modeling is at least five times smaller than statistical uncertainty.

\begin{figure}[thb]
  \centering
    \includegraphics[width=\cmsFigWidth]{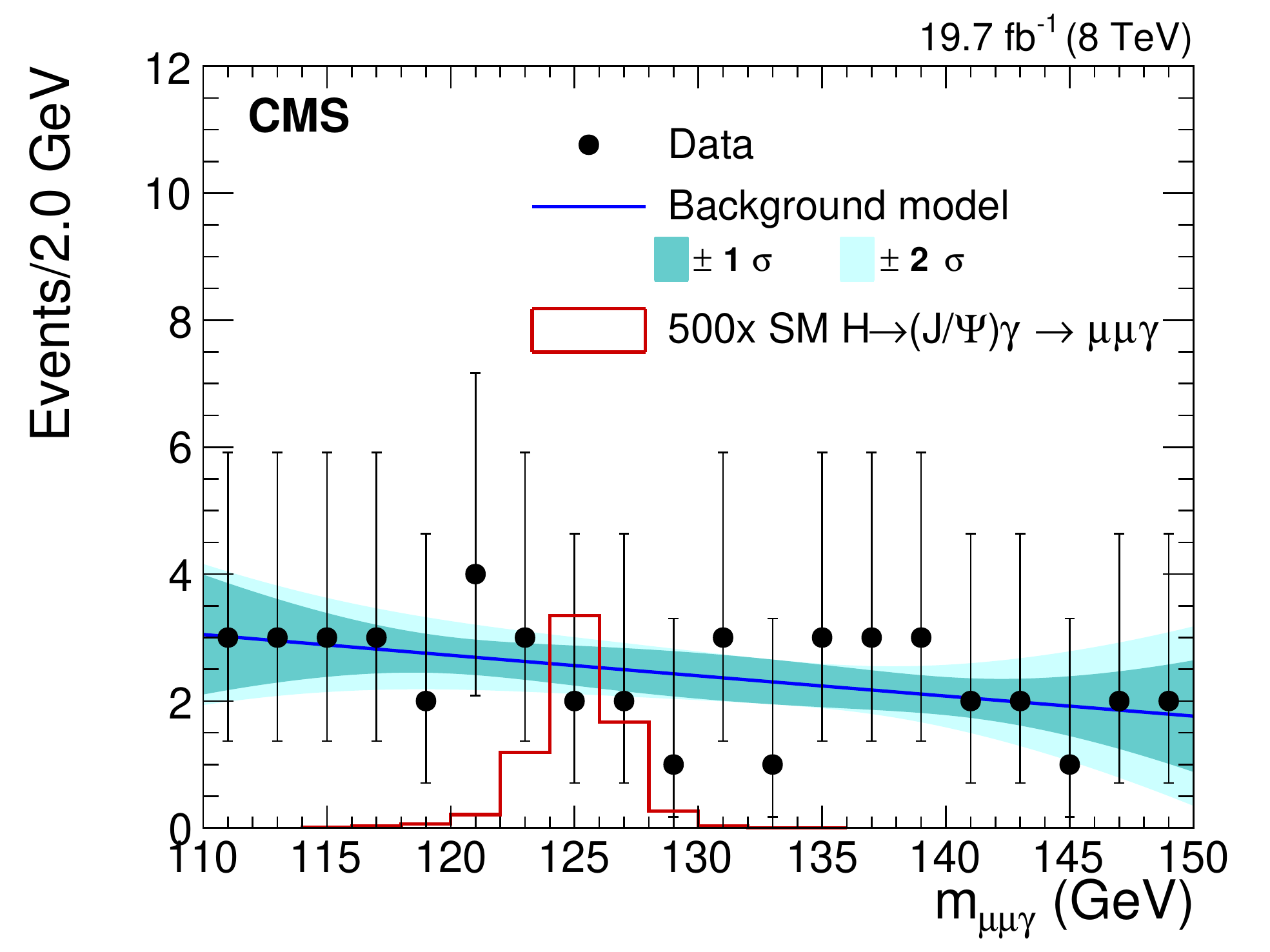}
    \caption{The $m_{\mu\mu\gamma}$ distribution for events with $2.9 < m_{\mu\mu} < 3.3\GeV$
      for 8\TeV data (points with error bars), together with the result of a
      background-only fit to the data. The 1$\sigma$ and 2$\sigma$ uncertainty bands
      represent the uncertainty in the parameters of the fitted function.
      The expected contribution from the $\PH\to(\JPsi)\gamma\to\mu\mu\gamma$ process of the SM H
      with $m_\PH=125\GeV$, scaled up by a factor of~500, is shown as a histogram.
      \label{fig:fit-jp}
    }
\end{figure}

For the $\PH\to(\JPsi)\gamma$ search, where only the single Higgs boson mass hypothesis
$m_\PH=125\GeV$ is investigated, a fit to a polynomial of second degree is performed over
the 110--150\GeV mass range (Fig.~\ref{fig:fit-jp}).

The signal model in all three cases is obtained from an unbinned fit to the mass
distribution of the corresponding sample of simulated events to a Crystal Ball
function~\cite{CB-Oreglia} plus a Gaussian function.

\section{Results}
\label{sec:results}
The data are used to derive upper limits on the Higgs boson cross section times branching
fraction, $\sigma(\Pp\Pp\to\PH)\,\mathcal{B}(\PH\to\gamma^*\gamma\to\ell\ell\gamma)$
divided by that expected for a SM Higgs boson, for $m_{\ell\ell} < 20\GeV$.  No
significant excess above background is observed in the full mass range,
$120<m_\PH<150\GeV$, with a maximum excess of less than two standard deviations.  In the
electron channel a correction is made to account for the events that are removed by the
requirement of $m_{\Pe\Pe} < 1.5\GeV$ due to the trigger and reconstruction inefficiencies
described above.

The exclusion limits are calculated using the modified frequentist CL$_\mathrm{s}$
method~\cite{cite:l1,cite:l2,cite:l3, Asympt, RooStats}.  An unbinned evaluation over the
full mass range of data is used.  The uncertainty in the limit is dominated by the size of
the data sample and systematic uncertainties have a small impact.

The systematic uncertainty in the limits results only from the uncertainty in the signal
description, as the background is obtained from data and biases in the fitting procedure
have been found to be negligible.  A summary of the systematic uncertainties is given in
Table~\ref{tab:syst}. The uncertainty can be separated into the uncertainty resulting from
theoretical predictions and from the uncertainty in detector reconstruction and selection
efficiency.

\begin{table}[h]
  \centering
    \topcaption{Systematic uncertainties affecting the signal}
    \label{tab:syst}
    \begin{tabular}{l|c}

      ~~~~~~~~ Source & Uncertainty \\
      \hline
      Integrated luminosity (ref.~\cite{CMS-PAS-LUM-13-001}) & 2.6\%\\
      \hline
      Theoretical uncertainties: &\\
      ~~~ PDF                                       & 2.6--7.5\%\\
      ~~~ Scale                                     & 0.2--7.9\%\\
      ~~~ $\PH\to\gamma^*\gamma\to\ell\ell\gamma$ branching fraction  &  10\% \\
      \hline
      Experimental uncertainties: &\\
      ~~~ Pileup reweighting                     & 0.8\%\\
      ~~~ Trigger efficiency, \PGm\,(\Pe) channel  & 4\,(2)\%\\
      ~~~ Muon reconstruction efficiency       & 11\%\\
      ~~~ Electron reconstruction efficiency  & 3.5\%\\
      ~~~ Photon reconstruction efficiency    & 0.6\%\\
      ~~~ $m_{\ell\ell\gamma}$ scale, \PGm\,(\Pe) channel & 0.1 (0.5)\%\\
      ~~~ $m_{\ell\ell\gamma}$ resolution, \PGm\,(\Pe) channel & 10 (10)\%\\
    \end{tabular}

\end{table}

Theoretical uncertainties come from the effects of the PDF choice on signal cross section,
the missing higher-order calculations
(scale)~\cite{cite:cs1,cite:cs2,cite:cs3,cite:cs4,cite:pdf}, and the uncertainty in the
prediction on the Higgs boson decay branching fraction~\cite{Heinemeyer:2013tqa,Passarino}.  The
uncertainty due to the muon reconstruction efficiency, 11\%, is obtained from data using
$\JPsi\to\mu\mu$ events. It is dominated by the statistical uncertainty of the data
sample. In the electron channel, the corresponding uncertainty, 3.5\%, is obtained from
simulation.  The 11\% uncertainty estimated for the muon identification efficiency is
sufficiently small and it has no impact on our result, thus no simulation study was
attempted, although it could greatly reduce the uncertainty.

\begin{figure}[thp]
  \centering \includegraphics[width=\cmsFigWidth]{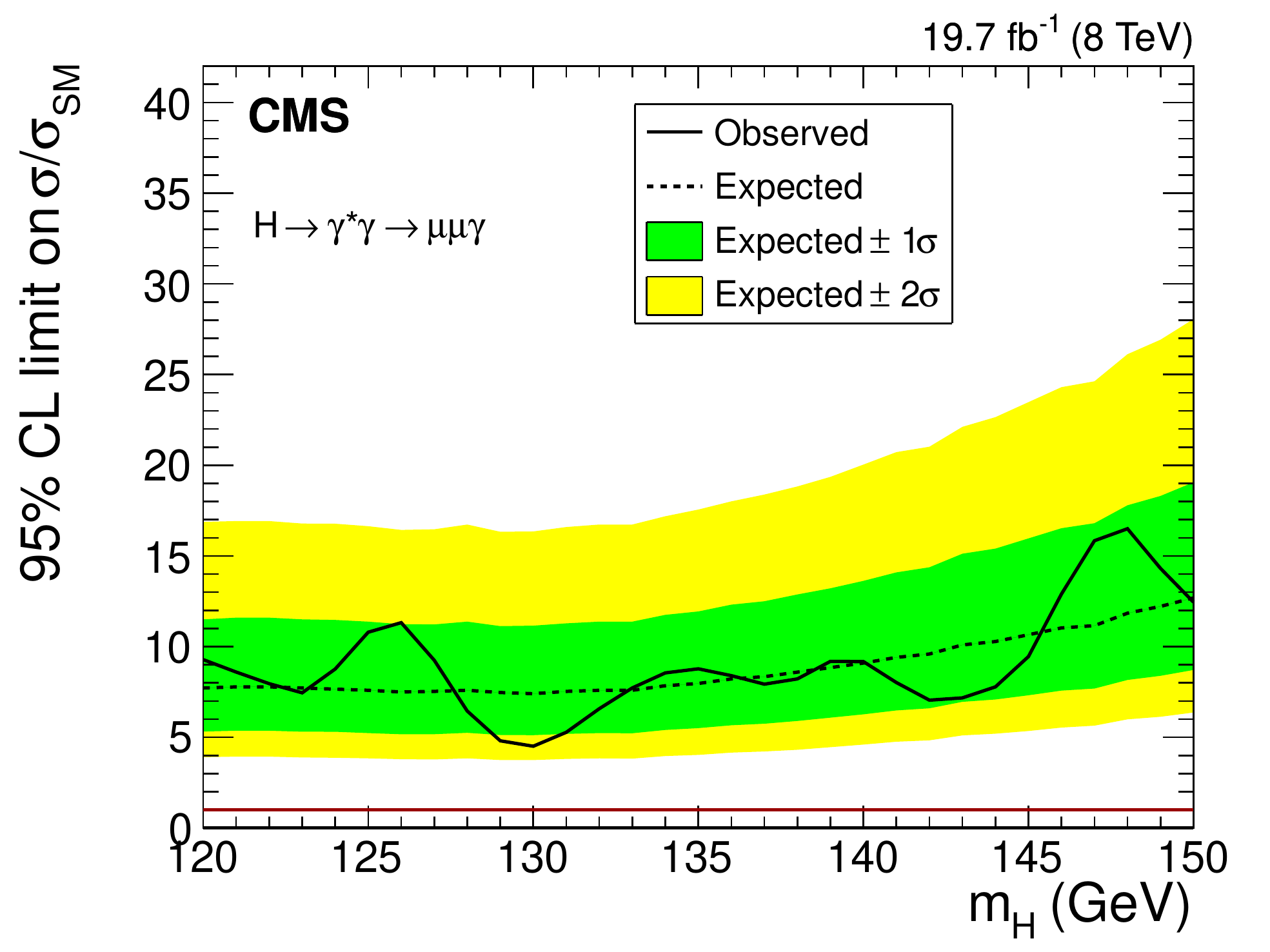} \includegraphics[width=\cmsFigWidth]{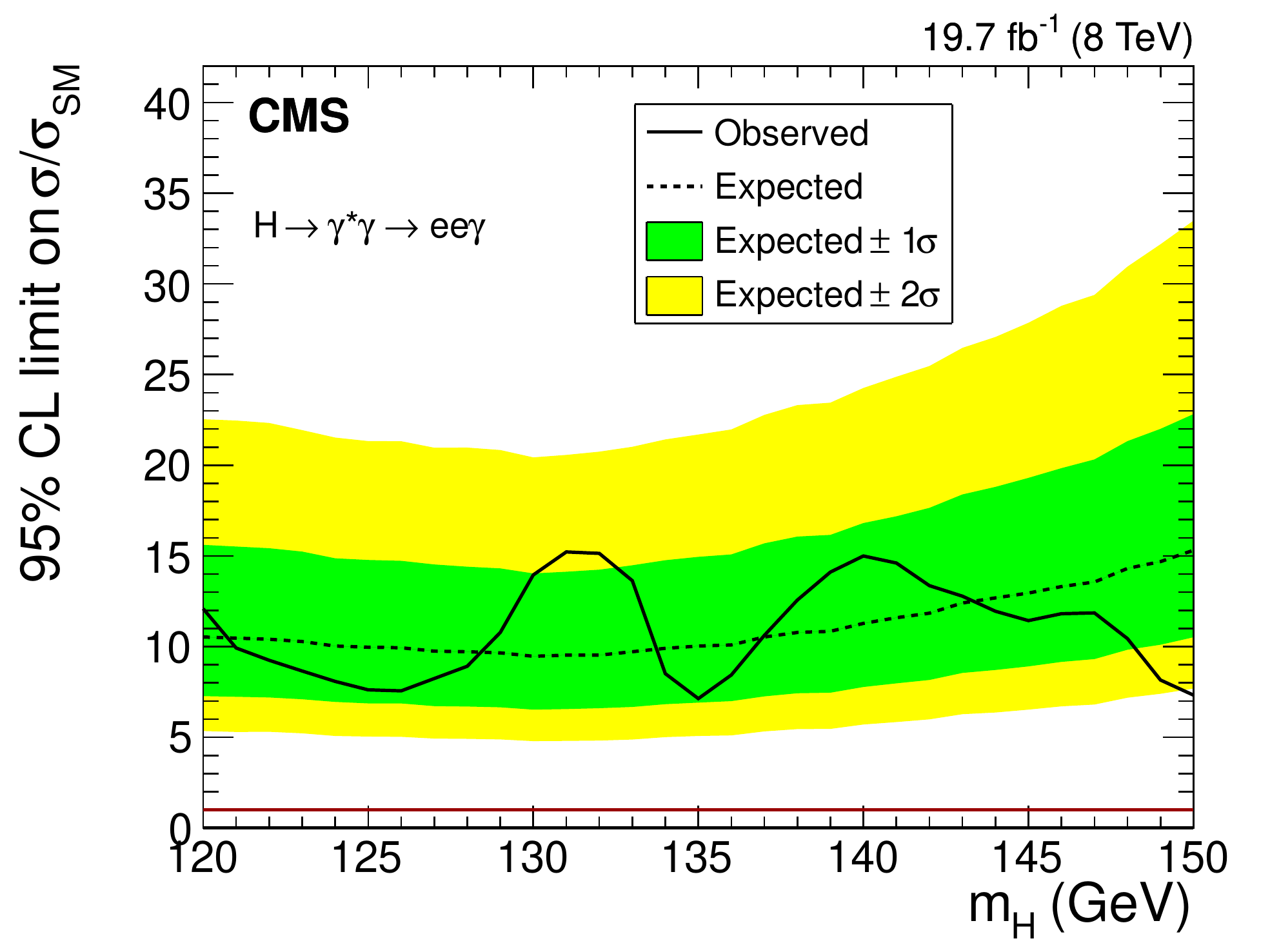} \includegraphics[width=\cmsFigWidthX]{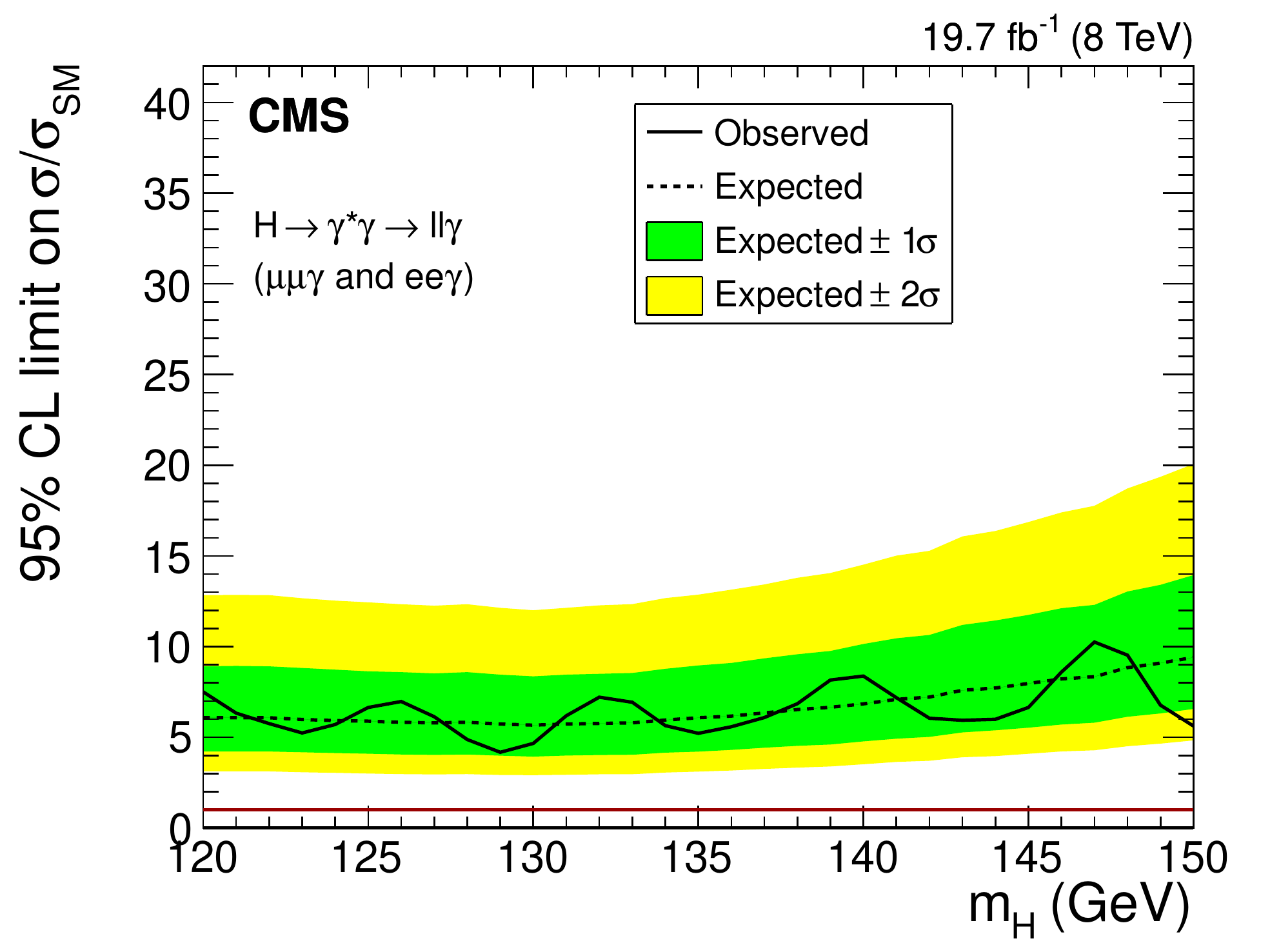} \caption{The
    95\% CL exclusion limit, as a function of the mass hypothesis,
    $m_\PH$, on $\sigma/\sigma_{SM}$, the cross section times the
    branching fraction of a Higgs boson decaying into a photon and a
    lepton pair with $m_{\ell\ell} < 20\GeV$, divided by the SM value.
    (\cmsUpper) muon, (\cmsMiddle) electron channels, (bottom) statistical
    combination of the results in the two
    channels \label{fig:limit-1}}.
\end{figure}

The expected and observed individual and combined $\mu\mu\gamma$ and $\Pe\Pe\gamma$ limits
are shown in Fig.~\ref{fig:limit-1}.  The limits are calculated at 1\GeV intervals in the
120--150\GeV mass range.  The median expected exclusion limits at 95\% confidence
level (CL) are between 6 and 10 times the SM prediction and the observed limit ranges
between about 5 and 11 times the SM.  The observed (expected) limit for $m_\PH = 125\GeV$
is 6.7 ($5.9^{+2.8}_{-1.8}$) times the SM prediction.

\begin{figure}[tbh]
  \centering
    \includegraphics[width=\cmsFigWidthX]{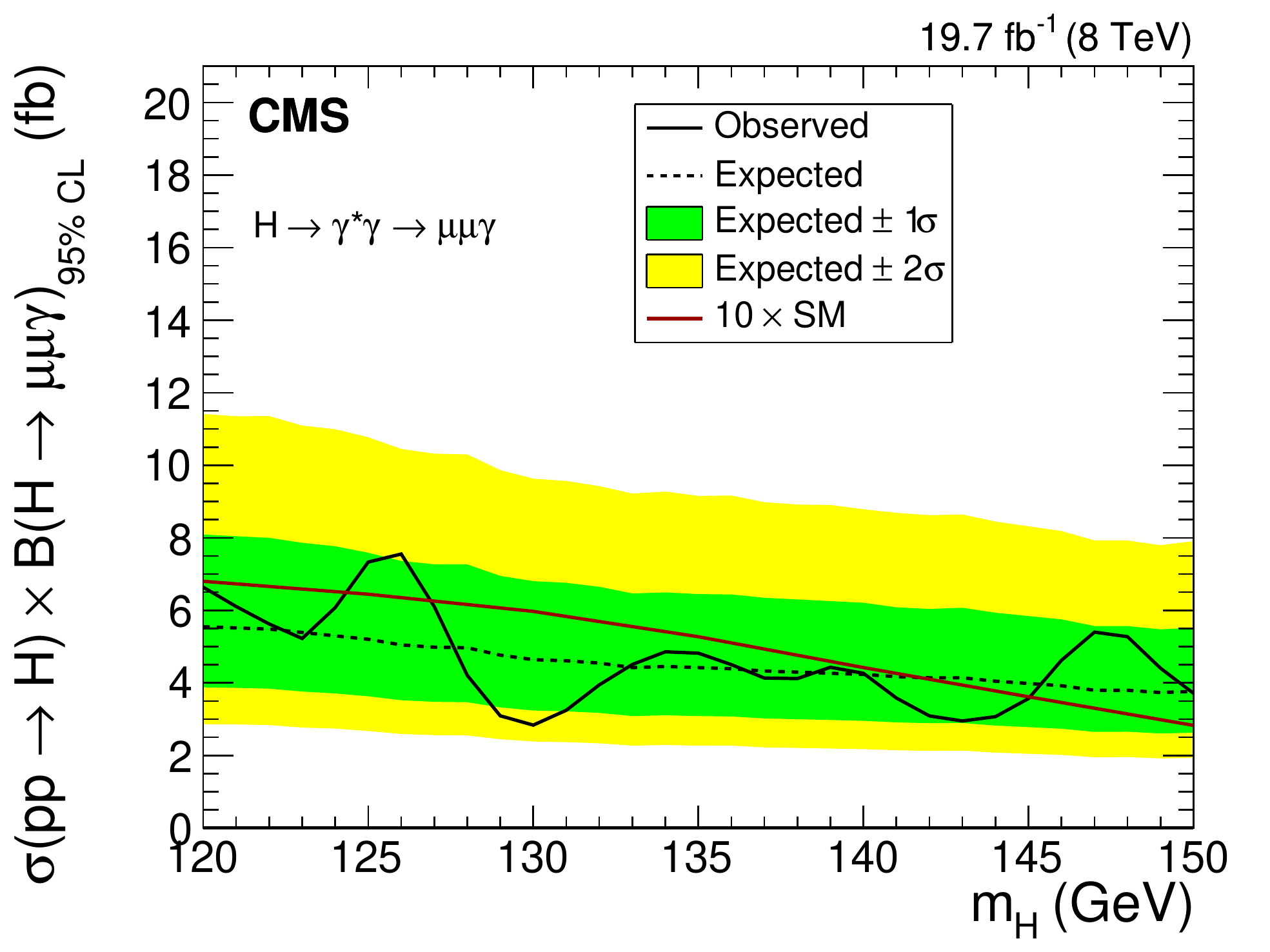}
    \caption{The 95\% CL exclusion limit on $\sigma(\Pp\Pp\to \PH)\,\mathcal{B}(\PH\to\mu\mu\gamma)$,
      with $m_{\mu\mu} < 20\GeV$, for a Higgs-like particle, as a function of the mass hypothesis, $m_\PH$.
      \label{fig:limit-xsBR}}

\end{figure}

The 95\% CL exclusion limits on $\sigma(\Pp\Pp\to \PH)\,
\mathcal{B}(\PH\to\mu\mu\gamma)$ for a narrow scalar particle without assuming the decay
kinematics of a SM Higgs boson, in the muon channel, are shown in
Fig.~\ref{fig:limit-xsBR}.  The observed (expected) limit for $m_\PH = 125\GeV$ is 7.3
($5.2^{+2.4}_{-1.6}$)\unit{fb}. The total signal efficiency is 24\% and almost independent of
the dimuon invariant mass. In the electron channel, however, this efficiency depends on
the dielectron mass, since it is strongly shaped by the selection.  For this reason the
corresponding limit in the electron channel is not evaluated.

Additionally, for the SM Higgs boson with $m_\PH=125\GeV$, we place an upper limit for a
$2.9 < m_{\ell\ell} < 3.3\GeV$ region in the muon channel: $\sigma(\Pp\Pp\to \PH)\,
\mathcal{B}(\PH\to\mu\mu\gamma) < 1.80$\unit{fb}, while the expected limit is
$1.90\pm0.97$\unit{fb}.  One can interpret this result as an upper limit on $\sigma(\Pp\Pp\to
\PH)\, \mathcal{B}(\PH\to (\JPsi)\gamma\to\mu\mu\gamma)$ and obtain for the branching
fraction, $\mathcal{B}(\PH\to(\JPsi)\gamma) < 1.5\times10^{-3}$ at 95\% CL, which is about
540 times the prediction in Ref.~\cite{hToJPsiGamma-2014}. The limit on the branching
fraction at 90\% CL is $\mathcal{B}(\PH\to(\JPsi)\gamma) < 1.2\times10^{-3}$. The number
of events present in this $m_{\mu\mu}$ mass window coming from the $\PH \to \gamma^*\gamma
\to \mu\mu\gamma$ is large compared to the $\PH\to (\JPsi)\gamma\to\mu\mu\gamma$ (as shown
in Table~\ref{tab:yield}). On the other hand it is small compared to the total background,
hence it is considered as a part of the background when extracting the limit on
$\mathcal{B}(\PH\to(\JPsi)\gamma)$.

\section{Summary}
\label{sec:Summary}

A search for a Higgs boson decay $\PH\to\gamma^*\gamma\to\ell\ell\gamma$ is presented.  No
excess above the background predictions has been found in the three-body invariant mass
range $120<m_{\ell\ell\gamma}<150\GeV$. Limits on the Higgs boson production cross section
times the $\PH\to\gamma^*\gamma\to\ell\ell\gamma$ branching fraction divided by the SM
values have been derived.  The observed limit for $m_\PH=125\GeV$ is about 6.7 times the
SM prediction.  Limits at 95\% CL on
$\sigma(\Pp\Pp\to \PH)\,\mathcal{B}(\PH\to\mu\mu\gamma)$ for a narrow resonance are also
obtained in the muon channel.  The observed limit for $m_\PH = 125\GeV$ is 7.3\unit{fb}.
Events consistent with the $\JPsi$ in dimuon invariant mass are used to set a 95\% CL
limit on the branching fraction $\mathcal{B}(\PH\to(\JPsi)\gamma) < 1.5\times10^{-3}$,
that is, 540 times the SM prediction for $m_\PH=125\GeV$.

\begin{acknowledgments}
We congratulate our colleagues in the CERN accelerator departments for the excellent performance of the LHC and thank the technical and administrative staffs at CERN and at other CMS institutes for their contributions to the success of the CMS effort. In addition, we gratefully acknowledge the computing centres and personnel of the Worldwide LHC Computing Grid for delivering so effectively the computing infrastructure essential to our analyses. Finally, we acknowledge the enduring support for the construction and operation of the LHC and the CMS detector provided by the following funding agencies: BMWFW and FWF (Austria); FNRS and FWO (Belgium); CNPq, CAPES, FAPERJ, and FAPESP (Brazil); MES (Bulgaria); CERN; CAS, MoST, and NSFC (China); COLCIENCIAS (Colombia); MSES and CSF (Croatia); RPF (Cyprus); MoER, ERC IUT and ERDF (Estonia); Academy of Finland, MEC, and HIP (Finland); CEA and CNRS/IN2P3 (France); BMBF, DFG, and HGF (Germany); GSRT (Greece); OTKA and NIH (Hungary); DAE and DST (India); IPM (Iran); SFI (Ireland); INFN (Italy); MSIP and NRF (Republic of Korea); LAS (Lithuania); MOE and UM (Malaysia); CINVESTAV, CONACYT, SEP, and UASLP-FAI (Mexico); MBIE (New Zealand); PAEC (Pakistan); MSHE and NSC (Poland); FCT (Portugal); JINR (Dubna); MON, RosAtom, RAS and RFBR (Russia); MESTD (Serbia); SEIDI and CPAN (Spain); Swiss Funding Agencies (Switzerland); MST (Taipei); ThEPCenter, IPST, STAR and NSTDA (Thailand); TUBITAK and TAEK (Turkey); NASU and SFFR (Ukraine); STFC (United Kingdom); DOE and NSF (USA).

Individuals have received support from the Marie-Curie programme and the European Research Council and EPLANET (European Union); the Leventis Foundation; the A. P. Sloan Foundation; the Alexander von Humboldt Foundation; the Belgian Federal Science Policy Office; the Fonds pour la Formation \`a la Recherche dans l'Industrie et dans l'Agriculture (FRIA-Belgium); the Agentschap voor Innovatie door Wetenschap en Technologie (IWT-Belgium); the Ministry of Education, Youth and Sports (MEYS) of the Czech Republic; the Council of Science and Industrial Research, India; the HOMING PLUS programme of the Foundation for Polish Science, cofinanced from European Union, Regional Development Fund; the Compagnia di San Paolo (Torino); the Consorzio per la Fisica (Trieste); MIUR project 20108T4XTM (Italy); the Thalis and Aristeia programmes cofinanced by EU-ESF and the Greek NSRF; the National Priorities Research Programme by Qatar National Research Fund; the Rachadapisek Sompot Fund for Postdoctoral Fellowship, Chulalongkorn University (Thailand); and the Welch Foundation.
\end{acknowledgments}

\bibliography{auto_generated}
\cleardoublepage \appendix\section{The CMS Collaboration \label{app:collab}}\begin{sloppypar}\hyphenpenalty=5000\widowpenalty=500\clubpenalty=5000\textbf{Yerevan Physics Institute,  Yerevan,  Armenia}\\*[0pt]
V.~Khachatryan, A.M.~Sirunyan, A.~Tumasyan
\vskip\cmsinstskip
\textbf{Institut f\"{u}r Hochenergiephysik der OeAW,  Wien,  Austria}\\*[0pt]
W.~Adam, E.~Asilar, T.~Bergauer, J.~Brandstetter, E.~Brondolin, M.~Dragicevic, J.~Er\"{o}, M.~Flechl, M.~Friedl, R.~Fr\"{u}hwirth\cmsAuthorMark{1}, V.M.~Ghete, C.~Hartl, N.~H\"{o}rmann, J.~Hrubec, M.~Jeitler\cmsAuthorMark{1}, V.~Kn\"{u}nz, A.~K\"{o}nig, M.~Krammer\cmsAuthorMark{1}, I.~Kr\"{a}tschmer, D.~Liko, T.~Matsushita, I.~Mikulec, D.~Rabady\cmsAuthorMark{2}, B.~Rahbaran, H.~Rohringer, J.~Schieck\cmsAuthorMark{1}, R.~Sch\"{o}fbeck, J.~Strauss, W.~Treberer-Treberspurg, W.~Waltenberger, C.-E.~Wulz\cmsAuthorMark{1}
% ...
\vskip\cmsinstskip
% ...
1:~~Also at Vienna University of Technology, Vienna, Austria\\
2:~~Also at CERN, European Organization for Nuclear Research, Geneva, Switzerland\\
% ...

\end{sloppypar}
\end{document}